\documentclass[twocolumn,aps,prd,showpacs,preprintnumbers,amsmath,amssymb,showpacs,showkeys,nofootinbib]{revtex4}
\usepackage{latexsym,epsfig,amsmath,amssymb}
\usepackage{amssymb}
\usepackage{amsmath}
\usepackage{amsfonts}
\usepackage{epsfig} 
\usepackage{verbatim} 

\newcommand{\seq}{\begin{subequations}}
\newcommand{\sen}{\end{subequations}}
\newcommand{\eq}{\begin{eqnarray}}
\newcommand{\en}{\end{eqnarray}}

\begin{document}

\title{New bounds on lepton flavor violating decays of vector mesons 
and the $Z^0$ boson} 

\noindent
\author{Thomas Gutsche$^1$, 
        Juan C. Helo$^{2}$,
        Sergey Kovalenko$^2$,        
        Valery E. Lyubovitskij$^1$\footnote{On leave of absence
        from Department of Physics, Tomsk State University,
        634050 Tomsk, Russia}
\vspace*{1.2\baselineskip}}

\affiliation{$^1$ Institut f\"ur Theoretische Physik,  
Universit\"at T\"ubingen,\\
Kepler Center for Astro and Particle Physics, \\  
Auf der Morgenstelle 14, D--72076 T\"ubingen, Germany
\vspace*{1.2\baselineskip} \\
$^2$ Departamento de F\'{i}sica, 
Universidad T\'{e}cnica Federico Santa Mar\'{i}a,\\
Centro Cient\'{i}fico-T\'ecnologico de  Valpara\'{i}so\\
Casilla 110-V, Valpara\'{i}so,
Valpara\'{i}so, Chile\\}

\date{\today}

\begin{abstract} 

We give an estimate for the upper bounds on rates of lepton flavor 
violating (LFV) decays $M\rightarrow \mu^{\pm} e^{\mp}$  of vector mesons 
$M = \rho^0, \omega, \phi, J/\psi, \Upsilon$ and the $Z^0$ boson 
in a model independent way, analyzing the corresponding lowest 
dimension effective operators. 
These operators also contribute to nuclear $\mu-e$-conversion.  
Based on this observation and using the existing experimental limits  
on this LFV nuclear process, we show that the studied two-body  
LFV decays of  vector bosons are strongly suppressed independent 
on the explicit realization of new physics. 
The upper limits on the rates of some of these decays are 
significantly more stringent than similar limits known in  
the literature. In view of these results experimental observation  
of the two-body LFV decays of vector bosons looks presently unrealistic. 

\end{abstract}



\maketitle

It is well known that  lepton flavor violation (LFV) is strongly 
suppressed in the Standard Model (SM) by a very small neutrino mass. 
Therefore, the observation of any LFV process would be a signal 
of physics beyond the SM.  Various avenues can be devised to study 
these issue. When elaborating on a strategy for searches of LFV 
these processes should be considered, which have the best prospects for 
discovery. This includes both the prospects for experimental identification 
of the LFV events and possible theoretical limitations on
the corresponding rates. 
Latter considerations may, for example, deal with model independent 
relations between different processes, some of which are already strongly 
limited by experimental data.  

In the present paper we study from this 
point of view LFV decays of vector mesons and the $Z^0$-boson
\begin{eqnarray}\label{proc-1} 
M\rightarrow \mu^{\pm} e^{\mp} \ \ \ \ \ \mbox{with} \ \ \ \ \  
M = \rho^0, \omega, \phi, J/\psi, \Upsilon, Z^{0} \,. 
\end{eqnarray}
The abundant production of  vector mesons and $Z^0$-bosons 
in current experiments naturally suggests to search for 
their two-body decays in the $\mu^{\pm}e^{\mp}$ 
final state, which is rather convenient for event identification. 
Recently the SND Collaboration at the 
BINP (Novosibirk)~\cite{Achasov:2009en} reported on the 
search for the LFV process $e^+e^-\to e\mu$ in the energy region 
$\sqrt{s}=984 - 1060$ MeV at the VEPP-2M $e^+e^-$ 
collider. They give a model independent upper limit on the 
$\phi\to e\mu$ branching fraction of
\begin{eqnarray}\label{SND-1}
{\rm Br}(\phi\to e\mu) < 2 \times 10^{-6}.
\end{eqnarray}
Also, there exist experimental limits for the $e\mu$ decay mode 
of $J/\psi$ and of the $Z^0$ boson~\cite{Nakamura:2010zz}: 
\begin{eqnarray}\label{PDG}
& &{\rm Br}(J/\psi \to e\mu)  < 1.1 \times 10^{-6}\,,\nonumber\\ 
& &{\rm Br}(Z^0 \to e\mu) < 1.7 \times 10^{-6} 
\end{eqnarray} 
and $\mu\tau$ decay mode of $\Upsilon$~\cite{Nakamura:2010zz}: 
\begin{eqnarray}\label{PDG2} 
& &{\rm Br}(\Upsilon \to \mu\tau) < 6.0 \times 10^{-6}\,. 
\end{eqnarray} 
In the near future this list may be extended by the results of 
other experimental collaborations. 
However, a natural question, which arises in this context, touches 
upon the prospects of this category of searches in view of possible 
theoretical limitations on the rates of these LFV decays.

In the literature there already exist stringent limits of this sort. 
For example, unitarity 
relations between the vector boson LFV decays given in~(\ref{proc-1}) 
and the pure leptonic LFV decay $\mu \to 3e$ have been exploited 
in Ref.~\cite{Nussinov:2000nm}. From the existing experimental 
bounds on the latter process the following stringent bounds were 
deduced~\cite{Nussinov:2000nm} 
\begin{eqnarray}\label{Nussinov}
& &{\rm Br}(\phi\to e\mu) \le 4 \times 10^{-17}\,, \nonumber\\ 
& &{\rm Br}(J/\psi \to e\mu)  \le 4 \times 10^{-13}\,, \nonumber\\ 
& &{\rm Br}(\Upsilon \to e\mu)  < 2 \times 10^{-9}\,, \\
 \nonumber 
& &{\rm Br}(Z^0 \to e\mu) < 5 \times 10^{-13}\,. 
\end{eqnarray}
In the present article we approach the LFV decays~(\ref{proc-1}) 
from another point of view relating these processes to nuclear $\mu-e$ 
conversion which is tightly constrained experimentally~\cite{Honecker:zf}. 

In Refs.~\cite{Faessler:2004jt,Faessler:2005hx} we studied the LFV process 
of nuclear $\mu^--e^-$ conversion in the framework of an effective 
Lagrangian approach without referring to any specific realization of 
physics beyond the SM responsible for LFV. 
We examined the impact of specific hadronization prescriptions on 
new physics contributions to nuclear $\mu^--e^-$ conversion and stressed
the importance of vector and scalar meson exchange between lepton and 
nucleon currents. In particular, we derived limits on various LFV couplings 
of vector mesons to the $\mu-e$ current using existing experimental data on 
$\mu^{-}-e^{-}$ conversion in nuclei. The purpose of the present paper is 
to use these limits to set upper bounds on the rates of the vector 
boson LFV decays given in~(\ref{proc-1}). In Ref.~\cite{Gutsche:2009vp} 
we already indicated in this framework upper limits on the rates of 
the LFV decays of $\rho^0$, $\omega$ and $\phi$ mesons. Here we extend our 
analysis to $J/\psi$, $\Upsilon$ and the $Z^0$ boson and compare our results 
with exiting experimental bounds~(\ref{SND-1}), (\ref{PDG}) and 
the theoretical predictions of Ref.~\cite{Nussinov:2000nm}. 

The contribution of vector bosons to $\mu^{-}-e^{-}$ conversion 
in nuclei is shown in~Fig.~1. In this diagram the upper vertex 
corresponds to the LFV interactions of vector bosons 
$M = \rho^0, \omega, \phi, J/\psi$, $\Upsilon$, $Z^0$ with $e, \mu$ 
given by  the following model--independent 
Lagrangian~\cite{Nussinov:2000nm,Faessler:2004jt}: 
\eq\label{eff-LV}
{\cal L}_{\rm eff}^{lM} &=& M^{\mu} 
\Big( \xi_V^{M} j_{\mu}^V\ + \xi_A^{M} j_{\mu}^A\  + {\rm h.c.} \Big) \, 
\,. 
\en 
Here the $\xi_{V,A}^{M}$ are effective vector and axial 
couplings of a vector boson $M$ to the LFV  lepton currents 
$j_{\mu}^V = \bar e \gamma_{\mu} \mu$ and  
$j_{\mu}^A = \bar e \gamma_{\mu} \gamma_{5} \mu$. 
The possible effect of additional non-minimal derivative couplings 
of vector bosons to the LFV lepton current will be considered later.

The lower vertex of the diagram in Fig.~1 is described by the 
nucleon-vector boson Lagrangian \cite{Gutsche:2009vp}:   
\eq\label{MN} 
{\cal L}_{MNN} = \frac{1}{2} \bar{N}\gamma^{\mu} N 
\sum\limits_M  g_{_{MNN}} M_\mu \,, 
\en 
where $g_{_{MNN}}$ are effective couplings. 
In this Lagrangian we neglected the derivative terms which are 
irrelevant for coherent $\mu^- - e^-$ conversion. 
The Lagrangian ${\cal L}_{MNN}$ is an extension of 
the conventional nucleon-vector meson 
Lagrangian~\cite{Weinberg:de,Mergell:1995bf,Kubis:2000zd}.  
In the case of the light $\rho^0$, $\omega$ and $\phi$ mesons 
we use values for $g_{_{MNN}}$ which are taken from an updated 
dispersive analysis~\cite{Mergell:1995bf,Meissner:1997qt} 
\begin{eqnarray}\label{VN-couplings}
g_{_{\rho NN}}= 4.0\,, \,\, 
g_{_{\omega NN}} = 41.8\,, \,\, 
g_{_{\phi NN}} = - 0.24\,. 
\end{eqnarray} 
\begin{figure}
\begin{center} 
\epsfig{file=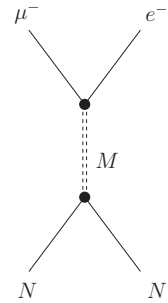, scale=.7} 
\end{center} 
\vspace*{-.75cm}
\caption{Contribution of intermediate vector bosons to nuclear 
$\mu^{-}-e^{-}$ conversion.} 
\end{figure} 
In addition we need an estimate for these effective couplings 
involving $J/\psi$, $\Upsilon$ and $Z^0$. 
The couplings $g_{_{J/\psi NN}}$ and $g_{_{\Upsilon NN}}$ can be extracted 
from data~\cite{Nakamura:2010zz} on $J/\psi \to N\bar N$ and 
$\Upsilon \to N\bar N$ decays: 
\eq 
& &\Gamma(J/\psi \to p\bar p) = 202 \pm 9 \ {\rm eV}\,, \nonumber\\
& &\Gamma(\Upsilon \to p\bar p) < 27 \pm 1 \ {\rm eV}.  
\en  
The two-body decay rate for $M \to p\bar p$, where $M = J/\psi$ or 
$\Upsilon$, is given in terms of the effective coupling constants 
$g_{_{MNN}}$ by: 
\begin{eqnarray}\label{M-dec}
\Gamma(M \to p \bar p) &=& 
\frac{g_{_{MNN}}^2 }{12\pi} \, m_M 
\biggl(1 - \frac{4m_N^2}{m_M^2}\biggr)^{1/2} \nonumber\\ 
&\times&\biggl(1 + \frac{2m_N^2}{m_M^2}\biggr) \,, 
\end{eqnarray}
Using the central value of $\Gamma(J/\psi \to p\bar p)$ and the 
upper limit for $\Gamma(\Upsilon \to p\bar p)$ \cite{Nakamura:2010zz} 
we get 
\begin{eqnarray}\label{J/psi-exp}
g_{_{J/\psi NN}} &\approx& 1.6 \times 10^{-3}\,, \label{gJNN} \\  
\label{Ups-exp}
g_{_{\Upsilon NN}} &<& 3.3 \times 10^{-4}  \label{gUNN_exp} \,. 
\end{eqnarray}
For our purposes the upper bound, as far the latter case, coupling 
is not sufficient. 
As it will be seen later from Eq.~(\ref{alpha-V-ex}), for the derivation 
of upper bounds on the effective couplings $\xi_{V,A}$ entering 
in~(\ref{eff-LV}) we need a definite estimate for the value of 
$g_{_{\Upsilon NN}}$. This can be done on the basis of a QCD analysis 
of exclusive processes of heavy quarkonia as 
$M \to p \bar p$~\cite{Brodsky:1981kj,Chernyak:1983ej}. 
The corresponding transition amplitudes are generated 
by three-gluon annihilation between the heavy and the light quarks. 
In this approach the following expression for the coupling
$g_{MNN}$ of a heavy quarkonium state $M$ of mass $m_{M}$ with the 
nucleon was derived as 
\eq 
g_{_{MNN}} = g \, \alpha_s^3(m_M^2) \, \frac{f_M}{m_M^5}\, .
\en 
Here $g = 95.4$ GeV$^4$ represents the loop integral over nucleon wave 
functions. Using data for the leptonic decay constants 
$f_{J/\psi} = 416.4$ MeV and 
$f_{\Upsilon} = 715.5$ MeV~\cite{Nakamura:2010zz} we get: 
\begin{eqnarray}\label{Jpsi-est}
g_{_{J/\psi NN}}   &=& 0.14 \ \alpha_s^3(m_{J/\psi}^2)\,, \\
\label{Ups-est}
g_{_{\Upsilon NN}} &=& 0.9 \times 10^{-3} \ \alpha_s^3(m_{\Upsilon}^2)\,.  
\end{eqnarray}  
Above value (\ref{Jpsi-est}) for the coupling $g_{_{J/\psi NN}}$  
coincides with  the value (\ref{J/psi-exp}) extracted from the experimental 
data for the strong coupling constant $\alpha_s(m_{J/\psi}^2) = 0.226$, 
which is in a good agreement with the world average of  
$\alpha_s(m_{J/\psi}^2) \simeq 0.26$~\cite{Bethke:2009jm}. 
This exercise can also be regarded as a consistency check of the 
approaches evaluated in Refs.~\cite{Brodsky:1981kj,Chernyak:1983ej}. 
Similarly, using the central value of the latest result for 
$\alpha_s(m_{\Upsilon}^2) = 0.184^{+0.015}_{-0.014}$~\cite{Brambilla:2007cz}  
extracted from radiative $\Upsilon$ decays we find from Eq.~(\ref{Ups-est}) 
the value of 
\eq\label{gUNN}     
g_{_{\Upsilon NN}} = 5.6 \times 10^{-6} \,. 
\en 
This result is significantly smaller than the upper bo\-und~(\ref{Ups-exp}) 
set by experiment (for a discussion see also Ref.~\cite{Baru:1992sn}).

The coupling of a $Z^0$ boson to nucleons is well known (see, for 
instance,~\cite{Commings}). Since the axial nucleon current does not 
contribute to the dominant coherent channel of nuclear $\mu-e$ conversion 
we only need the coupling to the vector nucleon current, which in view of 
Eq.~(\ref{MN}) we denote as $g_{ZNN}$. Neglecting a possible but small 
contribution of strange and heavy sea quarks in the nucleon it 
is given as~\cite{Commings} 
\eq\label{gZNN} 
g_{_{Z^0 NN}} = \frac{g}{2\cos\theta_W} 
(1 - 4 \sin^2\theta_W) \approx 0.31 \,. 
\en 
Here we used
the following values of the SM parameters: 
$M_W = 80.399 \ {\rm GeV}$\,,  
$\sin^2\theta_W = 0.2322$\,,  
$G_F = 1.16637~\times~10^{-5} {\rm GeV}^{-2}\,.$ 

Starting from the Lagrangian  
${\cal L}_{\rm eff}^{M} = {\cal L}_{\rm eff}^{lM} + {\cal L}_{MNN}$ 
of Eqs.~(\ref{eff-LV}) and (\ref{MN}) it is straightforward to derive 
the contribution of the diagram in Fig.~1 to the total $\mu^- - e^-$ 
conversion branching ratio \cite{Faessler:2004jt}. To the leading order  
in the non-relativistic reduction the coherent $\mu^--e^-$ conversion  
branching ratio takes the form~\cite{Kosmas:ch} 
\begin{equation} 
R_{\mu e^-}^{coh} \ = \  
\frac{{\cal Q}} {2 \pi } \  \   
\frac{p_e E_e } 
{ \Gamma ({\mu^-\to {\rm capture}}) } 
\, , 
\label{Rme}
\end{equation} 
where $p_e, E_e$ are 3-momentum and energy of the outgoing electron 
(for details see Ref.~\cite{Faessler:2004jt,Faessler:2005hx});
$\Gamma ({\mu^-\to {\rm capture}})$ is the total rate of the ordinary
muon capture reaction. The factor ${\cal Q}$ in Eq.~(\ref{Rme}) 
has the form~\cite{Faessler:2004jt,Kosmas:2001mv} 
\begin{eqnarray}
\nonumber
\hspace*{-1cm}
{\cal Q} &=& |({\cal M}_{p} + {\cal M}_{n})\alpha_{VV}^{(0)}+ 
({\cal M}_{p} - {\cal M}_{n})\alpha_{VV}^{(3)} |^2 +\\
&+&|({\cal M}_{p} + {\cal M}_{n})\alpha_{AV}^{(0)}+
({\cal M}_{p} - {\cal M}_{n})\alpha_{AV}^{(3)} |^2 \, . 
\label{Rme.1} 
\end{eqnarray} 
It contains the nuclear matrix elements ${\cal M}_{p,n}$ which have been
calculated numerically in Refs.~\cite{Kosmas:2001mv} for various nuclei. 
Here we consider $\mu^--e^-$ conversion in ${}^{48}$Ti studied by
the SINDRUM II Collaboration~\cite{Honecker:zf}. For this nucleus
we have ${\cal M}_{p}\approx 0.104, {\cal M}_{n}\approx 0.127$.
The  ${\cal Q}$ factor also contains the LFV lepton-nucleon parameters
$\alpha_{VV,AV}$.
For the contribution of the vector boson-exchange diagram in Fig.~1 
these coefficients are expressed in terms of the 
LFV couplings $\xi_{V,A}^{\rho,\omega,\phi}$ of Eq.~(\ref{eff-LV}) 
as~\cite{Faessler:2004jt,Faessler:2005hx,Gutsche:2009vp}
\begin{eqnarray} \label{alpha-V-ex}
\alpha_{aV}^{(3)} &=& - \frac{1}{2} 
\frac{g_{_{\rho NN}}}{m_{\rho}^2 
+ m_{\mu}^2} \xi_{a}^{\rho},
\\   
\nonumber
\alpha_{aV}^{(0)}&=& -\frac{1}{2} \sum\limits_H  
\frac{g_{_{H NN}}}{m_{H}^2 
+ m_{\mu}^2} \xi_{a}^{H} \,, 
\end{eqnarray}
where $a=V,A$; $H = \omega$, $\phi$, $J/\psi$, 
$\Upsilon$, $Z^0$;  
$m_M$ and $m_\mu$ the vector boson and muon masses, respectively. 

In Ref.~\cite{Faessler:2004jt} we extracted upper limits on the couplings 
$\alpha_{aV}^{(i)}$ from the experimental upper bounds on $\mu^{-}-e^{-}$ 
conversion in ${}^{48}$Ti reported by the SINDRUM II 
Collaboration~\cite{Honecker:zf}. These limits can be translated into 
bounds on the LFV couplings $\xi_{V,A}^{\rho,\omega,\phi}$. Assuming 
that no accidental cancellations occur between the different terms 
in~(\ref{alpha-V-ex}) and using the values of the vector boson-nucleon 
couplings  from Eq.~(\ref{VN-couplings}), (\ref{gJNN}), (\ref{gUNN}) 
and (\ref{gZNN}) we get 
for non-derivative couplings 
\begin{eqnarray}\label{limits-on-LFV-couplings}
& &\xi_{a}^{\rho} \leq 3.6 \times 10^{-12}\,, \quad  \xi_{a}^{\omega} 
\leq 3.6 \times 10^{-14}\,, \nonumber\\
& &\xi_{a}^{\phi} \leq 1.0 \times 10^{-11}\,, \quad
\xi_{a}^{J/\psi} \leq 1.4 \times 10^{-8}\,, \\ 
& &\xi_{a}^{\Upsilon} \leq 3.5 \times 10^{-5}\,, \quad  
\xi_{a}^{Z^0} \leq 6.4 \times 10^{-8}\,. \nonumber
\end{eqnarray}
Note that the Lagrangian (\ref{eff-LV}) also governs 
the LFV decay $M \to e \mu$ of vector bosons. Thus, using the limits 
from Eq.~(\ref{limits-on-LFV-couplings}) we can set upper bounds on 
the rates of these two-body decays. Their branching ratios are given by: 
\eq
{\rm Br}(M \to e \mu) \simeq 
\frac{(\xi_{V}^{M})^{2}+(\xi_{A}^{M})^{2}}{12\pi \Gamma^{M}_{\rm tot}} \, 
m_M  \, \biggl( 1 - \frac{3}{2} r_M^2 \biggr) \,, 
\en 
where $r_M = m_\mu/m_M$ and $\Gamma^{M}_{\rm tot}$ is the total decay width 
of boson $M$. Here we neglect the electron mass. With the limits set by 
Eq.~(\ref{limits-on-LFV-couplings}) we get the following  upper limits on 
the branching ratios of the vector boson LFV decays 
\begin{eqnarray}\label{Limits-on-branchings}
&&{\rm Br}(\rho^0 \to e \mu) \leq 3.5 \times 10^{-24}\,,\nonumber\\
&&{\rm Br}(\omega \to e \mu) \leq 6.2 \times 10^{-27}\,,\nonumber\\ 
&&{\rm Br}(\phi \to e \mu) \leq 1.3 \times 10^{-21}\,,\\ 
&&{\rm Br}(J/\psi \to e \mu) \leq 3.5 \times 10^{-13}\,,\nonumber\\
&&{\rm Br}(\Upsilon \to e \mu) \leq 3.9 \times 10^{-6}\,,\nonumber\\ 
&&{\rm Br}(Z^0 \to e \mu) \leq 8.0 \times 10^{-15}\,. \nonumber
\end{eqnarray} 
The limit for the $J/\psi \to e \mu$ mode is compatible with the 
corresponding number extracted from $\mu\rightarrow 3e$ as done in 
Ref.~\cite{Nussinov:2000nm} and shown in Eq.~(\ref{Nussinov}). 
For the cases of the $\phi \to e \mu$ and $Z^0 \to e \mu$ decay modes 
we obtain significantly more stringent upper limits than in 
Ref.~\cite{Nussinov:2000nm} while for $\Upsilon \to e \mu$ 
our limit considerably weaker. The existing experimental upper bounds on 
some of these rates listed in~(\ref{SND-1}) and (\ref{PDG}) are by far 
much weaker than the limits set by theory, both as in 
Ref.~\cite{Nussinov:2000nm} and as discussed here. 

Finally, let us comment on the effect of non-minimal derivative couplings 
of vector bosons to the LFV lepton current in the upper vertex of the 
diagram in Fig.~1. One could imagine a situation where the 
minimal non-derivative couplings in the effective Lagrangian~(\ref{eff-LV}) 
are substituted by the following derivative 
couplings~\cite{Nussinov:2000nm,Faessler:2004jt}: 
\eq\label{eff-LV-ext}
{\cal L}_{\rm eff}^{lM} &=& \frac{1}{m_{_{M}}} \, 
\Big( \xi_T^{M} j_{\mu\nu}^T\ + \xi_{\tilde T}^{M} j_{\mu\nu}^{\tilde T} 
\Big) 
\, M^{\mu\nu} + {\rm h.c.} \,,
\en 
where $M^{\mu\nu} = \partial^\mu M^\nu - \partial^\nu M^\mu$ is the 
stress tensor. Here the $\xi_{T,\tilde T}^{M}$ are effective tensor and 
pseudotensor couplings of the bosons $M$ to the LFV lepton currents 
$j_{\mu\nu}^T = \bar e \sigma_{\mu\nu} \mu$ and 
$j_{\mu\nu}^{\tilde T} = \bar e \sigma_{\mu\nu}\gamma_5 \mu$. 
As was noted in Ref.~\cite{Nussinov:2000nm}, the derivative 
couplings~(\ref{eff-LV-ext}) would lead to significant weakening of  
the bounds in~(\ref{Nussinov}). This happens because the 
contribution of the virtual vector bosons to $\mu\rightarrow 3e$ is 
reduced in comparison to the case of the non-derivative 
couplings~(\ref{eff-LV}). This is also true for nuclear 
$\mu-e$-conversion and the derived bounds of~(\ref{Limits-on-branchings}).  
These bounds would have to be divided by the following 
factors~\cite{Nussinov:2000nm}:
$
q^{2}/m_{_{M}}^{2}\approx m_{\mu}^{2}/(2 m_{_{M}}^{2}) = 
10^{-2}[\rho^{0}, \omega]; 5\times10^{-3}[\phi]; 5.7\times 10^{-4}[J/\psi];
6.1\times 10^{-5}[\Upsilon]. 
$
However, even with this  weakening the limits~(\ref{Limits-on-branchings}) 
still exclude experimental observation of the LFV decays of vector mesons 
$\rho^0, \omega, \phi, J/\psi, \Upsilon \rightarrow \mu^{\pm}e^{\mp}$ 
in the near future. 
The situation with derivative coupling of the $Z^0$ boson to the leptons 
is different from the case of the mesons and was studied 
in Ref.~\cite{FloresTlalpa:2001sp}. 
This coupling is induced by a $SU_L(2)\times U_Y(1)$-invariant 
effective operator of dimension higher than four.  
Due to the electroweak gauge invariance couplings both  
of the $Z^0$-boson and of the photon 
to the LFV lepton current are induced by the same operator. Therefore, 
$Z^0 \to e \mu$ can be constrained from the existing experimental data on 
$\mu \to e \gamma$ and the electron electric dipole moment.  
In this way a very stringent bound 
$Br(Z^{0}\to e \mu) < 10^{-23} - 10^{-22}$ was derived in 
Ref.~\cite{FloresTlalpa:2001sp}.  This is significantly more stringent 
than both our bound in Eq.~(\ref{Limits-on-branchings}) and the 
bound~(\ref{Nussinov}) derived in Ref.~\cite{Nussinov:2000nm}.

{\it In conclusion,}  we extracted from the experimental bounds on 
nuclear $\mu^{-}-e^{-}$ conversion  new upper limits on the LFV couplings 
of the vector mesons and the $Z^0$ boson to the $e-\mu$ lepton current.  
Then we applied these limits to deduce upper bounds on the branching 
ratios of LFV decays of vector mesons and the $Z$-boson. 
The obtained upper bounds are shown in Eq. (\ref{Limits-on-branchings}). 
Our bounds for the decays $\rho^0, \omega \to e  \mu$ are new. 
The bounds for $\phi\to e \mu$ are significantly more stringent than 
the corresponding bounds existing in the literature. 
This conclusion indicates, in particular, that the 
nuclear $\mu-e$- conversion is more sensitive probe of LFV than  
the decays of $Z^{0}$ boson and vector mesons. 
In the latter case this is true at least 
in the sector of the lepton interactions with 
the mesons made of quarks of the first and second generation. 
On the other hand searches for the LFV decays of $Z^{0}$ boson and mesons 
remain an important experimental effort since their observation at 
the rates above the limits (\ref{Limits-on-branchings}) would be 
a manifestation of  new LFV physics, which does not fit into the present 
analysis. In particular, it may imply a non-trivial mechanism of self 
cancellation of different terms in (\ref{Rme.1}), which we considered 
as unnatural. Then $\mu-e$-conversion rate could remain below the 
experimental bound ~\cite{Honecker:zf} while allowing large rates of 
the LFV decays of $Z^{0}$ boson and vector mesons.

\begin{acknowledgments}

This work was supported by the DFG under Contract No. FA67/31-2,  
by CONICYT (Chile) via
Centro-Cient\'\i fico-Tecnol\'{o}gico de Valpara\'\i so PBCT ACT-028
and by FONDECYT project 1100582. 
This research is also part of the European Community-Research 
Infrastructure Integrating Activity 
``Study of Strongly Interacting Matter'' (HadronPhysics2, Grant Agreement 
No. 227431), Federal Targeted Program "Scientific and scientific-pedagogical 
personnel of innovative Russia" Contract No. 02.740.11.0238. 

\end{acknowledgments}

\end{document}